\documentclass[10pt,twocolumn,twoside]{IEEEtran}
\usepackage{graphicx}          
\usepackage[fleqn]{amsmath}    
\usepackage{amsfonts}          
\usepackage{indentfirst}       
\usepackage{makecell}          
\usepackage{subfigure}

\usepackage[breaklinks,colorlinks,linkcolor=red,citecolor=blue,urlcolor=black]{hyperref} 

\usepackage{algorithm}
\usepackage{algorithmic}

\usepackage{cite}
\usepackage[numbers, sort&compress]{natbib} 
\setcitestyle{open={},close={}}             

\begin{document}
	\title{Distributed Estimation for Interconnected Systems with Arbitrary Coupling Structures}
	\author{Yuchen Zhang, 	
	    	Bo Chen, 
	  	    Li Yu, and 
	  	    Daniel W.C. Ho
		\thanks{
			Y. Zhang, B. Chen and L. Yu are with the Department of Automation, Zhejiang University of Technology, Hangzhou 310023, China.(email: YuchenZhang95@163.com, bchen@aliyun.com, lyu@zjut.edu.cn).
			
			D. W. C. Ho is with the Department of Mathematics, City University of Hong Kong, Hong Kong, 999077. (email: madaniel@cityu.edu.hk).
	}}
	
	\maketitle
	
	\begin{abstract} 
			
		This paper is concerned with the problem of distributed estimation for time-varying interconnected dynamic systems with arbitrary coupling structures. To guarantee the robustness of the designed estimators, novel distributed stability conditions are proposed with only local information and the information from neighbors. Then, simplified stability conditions which do not require timely exchange of neighbors' estimator gain information is further developed for systems with delayed communication. By merging these subsystem-level stability conditions and the optimization-based estimator gain design, the distributed, stable and optimal estimators are proposed. Quite notably, these optimization solutions can be easily obtained by standard software packages, and it is also shown that the designed estimators are scalable in the sense of adding or subtracting subsystems. Finally, an illustrative example is employed to show the effectiveness of the proposed methods.
	\end{abstract}
	
	
	\IEEEpeerreviewmaketitle
	
	\section{Introduction}
	With the development of communication and sensor technology, the scale of systems is consistently increasing as they are getting more and more connected. As early as the 1960s, the concept of interconnected systems had been proposed [\cite{c1}], and interconnected systems have received more and more attention in recent decades due to their wide applications in power systems [\cite{c2}], multi-robot systems [\cite{c3}], complex networks [\cite{c4, c5}], and biological networks [\cite{c6}]. Generally, interconnected systems are high-dimensional complex systems composed of numerous dispersed subsystems, which can be state-coupled with their neighboring subsystems. The increased complexity of interconnected systems in terms of both system topologies and dynamics has prevented traditional estimation approaches from achieving satisfactory performance [\cite{c7}]. This can be mainly attributed to the poor scalability of centralized structure in traditional approaches. Firstly, the spatial distribution of subsystems will lead to high communication burden and field deployment cost for centralized methods. Meanwhile, the centralized methods also suffer heavy computational burden with the increase of the dimensions of interconnected systems. In addition, the intricate coupling structures of interconnected systems are not exploited in centralized methods, making it necessary to re-ensure stability when adding or subtracting subsystems.
	Therefore, it is imperative to consider advanced estimation approaches for interconnected systems to guarantee the accuracy and the stability of estimators.
	
	Over the past several decades, different decentralized/distributed estimation approaches have been developed in the fields of multi-agent systems [\cite{c8, c9}], multi-sensor systems [\cite{c10, c11, c12}], and interconnected systems [\cite{c13, c14, c15, c16, c17, c18, c19, c20, c21, c22, c23}] to decrease communication overhead and computational complexity. In these approaches, local estimators are designed based on their own information and the information form their neighboring subsystems. However, the arbitrary couplings among subsystems impose more significant challenge to distributed analysis for interconnected systems, especially in terms of stability. For this reason, most of existing distributed estimation approaches for interconnected systems are based on special coupling structures or communication structures. With structural assumptions, the designed distributed estimators can provide better estimation performance and their stability can be ensured by local analysis. For example, the optimal locally unbiased filter was proposed in [\cite{c13}] with specific structure for information exchange, while the centralized and distributed moving horizon estimators were developed in [\cite{c14}] for sparse banded interconnected systems. The sparsity structure was also exploited to decompose interconnected systems into interconnected overlapping subsystems with coupled states that can be locally observed, then the distributed Kalman filter [\cite{c15}] and the consensus based decentralized estimator [\cite{c16}] were designed. Meanwhile, a sub-optimal distributed Kalman filtering problem was addressed in [\cite{c17}] for a class of sequentially interconnected systems. Note that it is difficult for most interconnected systems to transform into these structures. A hopeful idea to address distributed estimation problem without any structure constraints is to combine stability conditions and distributed estimator design methods. For instant, by adding constraints on stability conditions for general interconnected systems, distributed estimators with decoupling strategy were designed in [\cite{c18, c19, c20}] and a moving horizon estimator was proposed in [\cite{c21}] with the assumption of uncorrelated local estimation errors. Besides, the distributed estimators with plug-and-play fashion were developed in [\cite{c22, c23}] by exploiting the properties of infinity norm for small gain based stability conditions. However, how to design stable and distributed estimation methods based on local and neighboring information for general interconnected systems is still an open question.
	
	Since the 1960s, the stability problem for general interconnected systems has received a great deal of attention [\cite{c24, c25, c26}]. To the best of our knowledge, besides the centralized analysis of stability for overall systems, the stability analysis methods for general interconnected systems can be divided into three categories: 1) methods based on scalar or vector Lyapunov function [\cite{c1, c27, c28}]; 2) methods based on small gain theorem [\cite{c29, c30}]; 3) methods based on dissipativity theory [\cite{c31, c32}]. For the first category, the stability conditions involving $M$-matrices are derived by investigating the internal stability for both subsystems and the overall interconnected systems. Unfortunately, tests for $M$-matrices are successful only when the couplings among subsystems are weak. In contrast, the stability conditions for the second category are obtained by analyzing the input-output stability of subsystems, where the couplings are treated as input terms from neighboring subsystems. It also requires weak coupling conditions for small gain theorem based methods, but the results are less conservative and can lead to relatively simple design guideline [\cite{c24}]. Another kind of input-output stability results in the third category are based on the concept of dissipativity, which are not necessarily weak coupling conditions due to their centralized analysis. However, the above stability conditions are not fully distributed, which means the knowledge of the dynamics and the couplings from neighboring subsystems is not enough in the analysis process. How to develop these stability conditions into scalable distributed conditions is still challenging. One way to address this problem is to derive the distributed stability conditions by totally local analysis [\cite{c18, c19, c20, c21, c22, c23}], where the stability of subsystems is locally and sequentially analyzed. Nevertheless, this approach  is much more conservative than the centralized results, i.e., weak coupling conditions or structural assumptions are still required. Another promising idea is the subsystem-level analysis for centralized stability conditions by decomposing them into distributed ones. For example, the work in [\cite{c33}] focused on decomposing a centralized dissipativity condition into distributed dissipativity conditions of individual subsystems. Note that these conditions require huge communication burden to exchange message matrices among subsystems and cannot be generalized to time-varying interconnected systems.
	
    It should be pointed out that the distributed estimator designed in [\cite{c19}] only provides stability conditions with specific subsystem coupling structures, which are further interpreted as the directed acyclic graphs of couplings in [\cite{c20}]. As for general subsystem connection structures, the design of distributed estimators with subsystem-level stability conditions is still challenging, and has not yet been fully solved. Motivated by the above analysis, we shall investigate the distributed estimation problem for general time-varying interconnected systems. The main contributions of this paper can be summarized as follows:
	\begin{itemize}
		\item \textbf{Distributed stability analysis}. The distributed stability conditions, which only require subsystem-level knowledge of dynamics and couplings, are proposed for local estimators. Then, the effect of couplings on distributed conditions is discussed.
		\item \textbf{Distributed stability under delayed communication}. The simplified distributed stability conditions are proposed for time-varying interconnected systems with one-step communication delay. It is shown that the simplified conditions do not need real-time exchange of subsystems' gain information and can ease communication burden.
		\item \textbf{Distributed estimators design}. By combining the distributed stability conditions and the optimization-based estimator gain design, a recursive, stable and optimal estimators for time-varying noisy interconnected systems are proposed, where an upper bound of local estimation error covariance is minimized. The proposed estimators are fully distributed, that is, only based on local and neighboring information.
	\end{itemize}
	
	\emph{Notations}: Define $\mathbb{N}_l := \{1,2,...,l\}$, where $l$ is a natural number excluding zero, and denote the set of $n$-dimensional real vectors by $\mathbb{R}^n$. Give sets $A$ and $B$, $A \setminus B$ represents the set of all elements of A that are not in $B$, and
	$A \cap B$ is the intersection set of $A$ and $B$. The superscript `$\mathrm{T}$' represents the transpose, while the symmetric terms in a symmetric matrix are denoted by `$*$'. The inverse of the matrix $A$ is denoted by $A^{-1}$, and $\mathrm{Tr}(A)$ represents the trace of the matrix $A$. The identity matrix with appropriate dimensions is represented as `$I$', and the matrix with all zero elements is denoted by `$\mathbf{0}$'. The notation $X>(<)0$ is a positive definite (negative definite) matrix, and $X\ge(\le)0$ is a positive semi-definite (negative semi-definite) matrix. The notation $\mathrm{col}\{a_1, ..., a_n\}$ means a column vector whose elements are $a_1, ..., a_n$, while $\mathrm{diag}\{\cdot\}$ stands for a block diagonal matrix. The mathematical expectation is denoted by $\mathrm{E}\{ \cdot \}$, and $\|A\|_2$ is the 2-norm of matrix $A$. Given a block matrix $A=[A_{i,j}]_{i \in \mathbb{N}_n, j\in \mathbb{N}_m}$, $A_{i,j}$ represents the $(i,j)$th block. The maximum eigenvalue of matrix $A$ is represented as $\lambda_{\mathrm{max}}(A)$.
	
	
	
	
	\section{Problem Formulation}
	\subsection{Time-varying Interconnected System Model}
	Consider a time-varying interconnected system $\mathbf{S}$ constructed by $l$ subsystems, where the state and measurement dynamics of the $i$th subsystem $\mathbf{S}_i, \ i \in \mathbb{N}_l$ is described as follows:
	\begin{equation}
	\label{E1}
	\mathbf{S}_i:
	\begin{cases}
	& \!\!\!\!\!\!
	\begin{aligned} 
	x_i(k+1) =
	& A_i(k) x_i(k) + \Gamma_i(k) w_i(k) \\
	& + \sum_{i_{\kappa}^\rho \in \Omega_i} A_{i, i_{\kappa}^\rho}(k) x_{i_{\kappa}^\rho}(k)
	\end{aligned}
	\\
	& \!\!\!\!\!\!
	y_i(k) = C_i(k) x_i(k) + D_i(k) v_i(k)
	\end{cases} \ \ \ \ i \in \mathbb{N}_l
	\end{equation}
	The vectors $x_i(k)\in \mathbb{R}^{n_i}$ and $y_i(k) \in \mathbb{R}^{m_i}$ denote the state and the measurement of the subsystem $\mathbf{S}_i$, respectively. Moreover, $A_i(k)$, $\Gamma_i(k)$, $A_{i,i_{\kappa}^\rho}(k)$, $C_i(k)$ and $D_i(k)$ are bounded matrices with appropriate dimensions, while the system noise $w_i(k)$ and the measurement noise $v_i(k)$ are uncorrelated Gaussian white noises satisfying
	\begin{equation}
	\label{E7}
	\begin{cases} 
	& \! \! \! \! \! 
	\mathrm{E} \left[ w_i(k) w_j(k_1) \right] = \delta_{i,j} \delta_{k,k_1} Q_{w_i}  \\
	& \! \! \! \! \!
	\mathrm{E} \left[ v_i(k) v_j(k_1) \right] = \delta_{i,j} \delta_{k,k_1} Q_{v_i}  \\
	& \! \! \! \! \! 
	\mathrm{E} \left[ w_i(k) v_j(k_1) \right] = 0 (\forall i,j,k,k_1)
	\end{cases}
	\end{equation}
	where $Q_{w_i}$ and $Q_{v_i}$ are the known covariances of $w_i(k)$ and $v_i(k)$, respectively. $\delta_{k,k_1}=0$ if $k \neq k_1$ and $\delta_{k,k_1}=1$ otherwise. The set of neighbors for subsystem $\mathbf{S}_i$ is denoted by $\Omega_i$, and the number of elements is $\theta_i \ (\theta_i <l)$. Therefore, the set $\Omega_i$ can be described as
	\begin{equation}
	\label{E2}
	\Omega_i = \{ i_{1}^\rho, ..., i_{\kappa}^\rho, ..., i_{\theta_i}^\rho\}
	\end{equation}
	The coupling structure of the system is determined by whether the matrix $A_{i, i_{\kappa}^\rho}(k)$ is a null matrix. Since there is no constraints on the spatial distribution of subsystems, the coupling structure can be arbitrary. Then, the following subset $\Sigma_i$ is defined:
	\begin{equation}
	\label{E3}
	\Sigma_i : = \{i_{\kappa}^\sigma \mid i_{\kappa}^\sigma \in \Omega_i \setminus \mathbb{N}_i \}
	\end{equation}
	where the number of elements for $\Sigma_i$ is $\xi_i \ (\xi_i \le \theta_i)$. \\
	
	\noindent \textbf{Remark 1.} Compared with the work in [\cite{c13, c14, c15, c16, c17}], the addressed interconnected system model in this paper does not require any structural assumptions (i.e., the sparsity assumption on couplings). In this case, the model in (\ref{E1}) is more general and can cover a large part of practical situations. For example, the heavy duty vehicle systems [\cite{c34}] with aerodynamic interconnections can be modeled as interconnected systems with strongly connected topologies in the form of (\ref{E1}). On the other hand, there is no constraint on the coupling strength for the interconnected system model in this paper, which is different from the work in [\cite{c22, c23}]. In other words, the upper bound of $\|A_{i,j}(k)\|_2$ can be arbitrarily large. However, the analysis of distributed stability and the distributed estimation problem for general interconnected system without any weak coupling assumptions will be more challenging. \\
	
	To collaboratively achieve system tasks, subsystems need to exchange their information via communication networks. Therefore, the distributed communication structure in the following assumption is required. \\	
	
	\noindent \textbf{Assumption 1 (Communication).} Each subsystem can communicate with its neighbors.\\	
	
	\noindent \textbf{Remark 2.} Notice that the communication structure in Assumption 1 is distributed and has a limited range of information broadcast due to the limitation on network bandwidth and energy constraints for subsystems. Unlike the centralized communication structure with one subsystem communicates with all the other subsystems, the considered distributed communication structure is more practical. On the other hand, we restrict ourselves to the time-varying interconnected systems with constantly varying dynamics and couplings due to its wider applications. Take blocked power systems as an example, the couplings among different blocks are changing with the real-time power dispatching [\cite{c35}]. For time-varying interconnected systems, the distributed stability conditions in [\cite{c33}] are not suitable anymore, and thus novel distributed stability analysis approaches are required.
	
		\subsection{Problem of Interest}
		\begin{figure*}[h]
			\begin{center}
				\includegraphics[height=5.7cm, width=16cm]{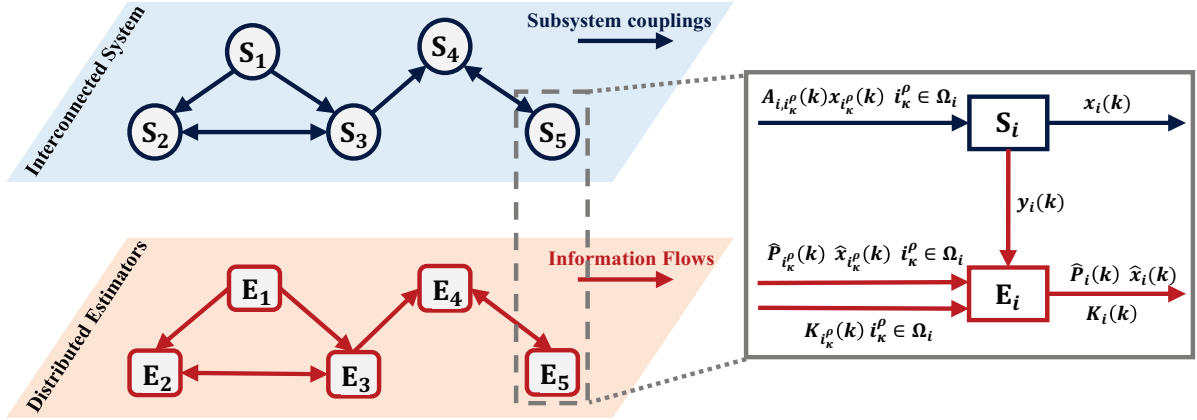}    
				\caption{An example of the structure for interconnected systems and distributed estimators.}  
				\label{fig1}                                 
			\end{center}                                 
		\end{figure*}
		The structure of distributed estimators for interconnected systems with local information flows is depicted in Fig. \ref{fig1}. It is assumed that subsystems can only know their own dynamics, and thus the local measurements and the estimates form neighbors are used for state reconstruction. The estimator $\mathbf{E}_i$ for the $i$th subsystem is proposed as
		\begin{equation}
		\label{E8}
		\mathbf{E}_i:
		\begin{cases}
		& \! \! \! \! \! \!
		\hat x^p_i (k) = A_i(k-1) \hat x_i(k-1) \\
		& \ \ \ \ \ \ \ \ 
		+ \sum_{i_{\kappa}^{\rho} \in \Omega_i} A_{i, i_{\kappa}^{\rho}}(k-1) \hat x_{i_{\kappa}^{\rho}}(k-1) 
		\\
		& \! \! \! \! \! \!
		\hat x_i(k) = \hat x^p_i(k) + K_i(k) \left[ y_i(k) - C_i(k) \hat x^p_i(k) \right]
		\end{cases}
		\end{equation}
		where $\hat x_i^p(k)$ and $\hat x_i(k)$ are the one-step prediction and the estimate of subsystem state $x_i(k)$, respectively. Then, the estimation error iteration for the $i$th subsystem is calculated by (\ref{E1}) and (\ref{E8}) as
		\begin{equation}
		\label{E9}
		\begin{cases}
		& \! \! \! \! \! \! 
		\tilde x_i^p(k) = A_i(k-1) \tilde x_i(k-1) + \Gamma_i(k-1) w_i(k-1) \\
		& \ \ \ \ \ \ \ \
		+ \sum_{i_{\kappa}^{\rho} \in \Omega_i} A_{i, i_{\kappa}^{\rho}}(k-1) \tilde x_{i_{\kappa}^{\rho}}(k-1) \\
		& \! \! \! \! \! \! 
		\tilde x_i(k) = K_{C_i}(k) \tilde x_i^p(k) - K_i(k) D_i(k) v_i(k)
		\end{cases}
		\end{equation}
		where $K_{C_i}(k) : = I-K_i(k)C_i(k)$, while $\tilde x_i^p(k)$ and $\tilde x_i(k)$ are the one-step prediction error and the estimation error, respectively. The one-step prediction error covariance $P_i^p(k) : = \mathrm{E}\left\{\tilde x_i^p(k) \left[ \tilde x_{i}^p(k) \right]^{\mathrm{T}} \right\}$ and the estimation error covariance $P_i(k) : = \mathrm{E}\left\{\tilde x_i(k) \tilde x_{i}^{\mathrm{T}}(k) \right\}$ can be calculated as	
		\begin{equation}
		\label{E50}
		\begin{cases}
		& \! \! \! \! \! \! 
		\begin{aligned}
		& P_i^p(k) = A_i(k-1) P_i(k-1) A_i^{\mathrm{T}}(k-1)  \\
		& \ \ + \Gamma_i(k-1) Q_{w_i} \Gamma_i^{\mathrm{T}}(k-1)  \\
		& \ \ + \sum_{i_{\kappa}^{\rho} \in \Omega_i} A_i(k-1) P_{i,i_{\kappa}^{\rho}}(k-1) A_{i,i_{\kappa}^{\rho}}^{\mathrm{T}}(k-1) \\
		& \ \ + \sum_{i_{\kappa}^{\rho} \in \Omega_i} A_{i,i_{\kappa}^{\rho}}(k-1) P_{i_{\kappa}^{\rho},i}(k-1) A_i^{\mathrm{T}}(k-1)  \\
		& \ \ + \sum_{i_{\kappa_1}^{\rho} \in \Omega_i} \sum_{i_{\kappa_2}^{\rho} \in \Omega_i} \left\{ A_{i,i_{\kappa_1}^{\rho}}(k-1) \right. \\
		& \left.  \ \ \ \ \ \  P_{i_{\kappa_1}^{\rho},i_{\kappa_2}^{\rho}}(k-1) A_{i,i_{\kappa_2}^{\rho}}^{\mathrm{T}}(k-1) \right\}
		\end{aligned}
		\\
		& \! \! \! \! \! \!
		\begin{aligned}
		P_i(k) = 
		& K_{C_i}(k) P_i^p(k) \left[ K_{C_i}(k) \right]^{\mathrm{T}}  \\
		& +  K_i(k) D_i(k) Q_{v_i} D_i^{\mathrm{T}}(k) K_i^{\mathrm{T}}(k)
		\end{aligned}
		\end{cases}
		\end{equation}
		The major concern of the distributed estimation problem is to design suitable gain matrices $K_i(k) \ (i \in \mathbb{N}_l)$ such that the estimation error is stable and the estimation performance index $J_i(k)$ is minimized. Specifically, the following definition is introduced to describe the property of stability for local estimators.\\

		\noindent \textbf{Definition 1 (Mean-square uniformly bounded).} For the interconnected system in (\ref{E1}), the proposed estimator (\ref{E8}) is mean-square uniformly bounded if for arbitrarily large $\delta_{p_i^0}$, there is $\delta_{p_i}(\delta_{p_i^0})>0$ (independent of $k_0$) such that
		\begin{equation}
		\label{E15}
		\| P_i(k_0) \|_2 \le \delta_{p_i^0} \Rightarrow \| P_i(k) \|_2 \le \delta_{p_i}
		\end{equation}
		
		\noindent However, it is usually difficult for subsystems to timely obtain the cross-covariances $P_{i,j}(k) : = \mathrm{E}\{ \tilde x_i(k) \tilde x^{\mathrm{T}}_j(k)\}$ by only local communication. Therefore, an upper bound of the estimation error covariance $\hat P_i(k) \ge P_i(k)$ is used instead and the performance index for local estimation is designed as $J_i(k) = \mathrm{Tr}\{ \hat P_i(k) \}$. 
		Here, the optimal estimator gain design for subsystems can be formulated as an optimization problem:
		\begin{equation}
		\label{E70}
		\begin{aligned}
		& \min_{K_i(k)} \mathrm{Tr}\{ \hat P_i(k) \} \\
		& \mathrm{s.t.} \ \ \hat P_i(k) \ge P_i(k) \ \ \mathrm{and} \ \ K_i(k) \in \mathcal{K}_i(k)
		\end{aligned}
		\end{equation}
		where $\mathcal{K}_i(k)$ is a subspace of stable estimator gains for subsystem $\mathbf{S}_i$ at the instant $k$. 
		
		In what follows, the augmented system dynamics and the augmented estimator iteration will be presented. By defining $x(k) : = \mathrm{col}\{x_1(k), ..., x_l(k)\} \in \mathbb{R}^{n}$, we can obtain the overall system dynamics as
		\begin{equation}
		\label{E4}
		\mathbf{S}:
		\begin{cases}
		& \!\!\!\!\!\! 
		x(k+1) = A(k) x(k) + \Gamma(k) w(k)
		\\
		& \!\!\!\!\!\!
		y(k) = C(k) x(k) + D(k) v(k)
		\end{cases}
		\end{equation}
		where $A(k) : =[A_{i,j}(k)]_{i,j \in \mathbb{N}_l}$ with $A_{i,i}(k)=A_i(k)$ and
		\begin{equation}
		\label{E5}
		\begin{cases}
		& \! \! \! \! \! \!
		y(k) : = \mathrm{col}\{y_1(k), ..., y_l(k)\} \\
		& \! \! \! \! \! \!
		\Gamma(k) : = \mathrm{diag}\{\Gamma_1(k), ..., \Gamma_l(k)\} \\
		& \! \! \! \! \! \!
		C(k) : = \mathrm{diag}\{C_1(k), ..., C_l(k) \} \\
		& \! \! \! \! \! \!
		D(k) : = \mathrm{diag}\{D_1(k), ..., D_l(k) \} \\
		& \! \! \! \! \! \!
		v(k) : = \mathrm{col}\{v_1(k), ..., v_l(k)\} \\
		& \! \! \! \! \! \!
		w(k) : = \mathrm{col}\{w_1(k), ..., w_l(k)\}
		\end{cases}
		\end{equation}
		The upper bounds of bounded matrices are $\|A(k)\|_2 \le \delta_a$, $\|\Gamma(k)\|_2 \le \delta_{\gamma}$, $\|C(k)\|_2 \le \delta_{c}$, $\|D(k)\|_2 \le \delta_{d}$, $\|A_{i,j}(k)\|_2 \le \alpha_{i,j}$ and $\|A_{i}(k)\|_2 \le \alpha_{i}$, respectively. Then, let us denote $\hat x(k) : = \mathrm{col}\{\hat x_1(k), ..., \hat x_l(k)\}$ and $\tilde x(k) : = \mathrm{col}\{\tilde x_1(k), ..., \tilde x_l(k)\}$. The following augmented estimator and the augmented estimation error iteration are obtained:
		\begin{equation}
		\label{E10}
		\begin{cases}
		& \! \! \! \! \! \!
		\begin{aligned}
		\hat x(k) = 
		& A(k-1) \hat x(k-1) \\
		& + K(k) \left[ y(k) - C(k) A(k-1) \hat x(k-1) \right] 
		\end{aligned}
		\\
		& \! \! \! \! \! \!
		\begin{aligned}
		\tilde x(k) = 
		& K_C(k) A(k-1) \tilde x(k-1) - K(k) D(k) v(k)   \\ 
		& + K_C(k) \Gamma(k-1) w(k-1)
		\end{aligned}
		\end{cases}
		\end{equation}
		where $K(k) : = \mathrm{diag}\{K_1(k), ..., K_l(k)\}$ and $K_C(k) : = I-K(k)C(k)$.
		
		Note that the stability of each local estimator depends on the stability of its neighboring estimators due to the interconnected estimation error $\tilde x_{i_{\kappa}^{\rho}}(k-1)$. Hence, it is difficult to determine $\mathcal{K}_i(k)$ and design a stable estimator in a totally local analysis. On the other hand, a totally centralized analysis for the augmented estimation error system needs the knowledge of dynamics and couplings from the overall system, which cannot apply to large-scale interconnected systems.
		
		Consequently, the aim of this paper is to address the following problems:
		\begin{itemize}
			\item[1)] \textbf{Distributed stability conditions analysis}: Analyze the distributed stability conditions such that the proposed estimator is mean-square uniformly bounded, where only subsystem-level knowledge of dynamics and couplings is required for each estimator.
			\item[2)] \textbf{Distributed estimator design}: Design distributed, stable and optimal estimators for time-varying interconnected systems with arbitrary coupling structures, where an upper bound of local estimation error covariance is minimized.\\
		\end{itemize}
		
		\noindent \textbf{Remark 3.} To design a fully distributed estimator, both the iteration form and the stability conditions for the estimator need to achieve local communication, computation and storage. Though the estimator in (\ref{E8}) only uses the information of local measurement and neighboring estimates, the local estimation errors are still interconnected. Therefore, the major difficulty for the distributed estimator design for general interconnected systems is to calculate the optimal estimator gain and maintain the stability without any globally interconnection information of estimation errors.		
		
	\section{Main Results}
	In this section, we firstly present distributed conditions to guarantee the stability of local estimators. Then, a fully distributed estimation approach is proposed by merging optimal and stable estimator gain designs.
	
	\subsection{Distributed Stability Conditions}	
	
	Let us denote the augmented estimation error covariance as $P(k):=\mathrm{E}\{\tilde x(k) \tilde x^{\mathrm{T}}(k)\}$ and it is calculated by
	\begin{equation}
	\label{E13}
	\begin{aligned}
	P(k) =
	& K_C(k) A(k-1) P(k-1) A^{\mathrm{T}}(k-1) K_C^{\mathrm{T}}(k) \\
	& + K_C(k) \Gamma(k-1) Q_w \Gamma^{\mathrm{T}}(k-1)   K_C^{\mathrm{T}}(k) \\
	& + K(k) D(k) Q_v D^{\mathrm{T}}(k) K^{\mathrm{T}}(k)
	\end{aligned}
	\end{equation} 
	where the matrices $Q_w : = \mathrm{diag}\{Q_{w_1}, ..., Q_{w_l}\}$ and $Q_v : = \mathrm{diag}\{Q_{v_1}, ..., Q_{v_l}\}$ are the augmented noise covariances. Then, the centralized stability conditions are derived by the following proposition. Its proof appears in the Appendix. \\
	
	\noindent \textbf{Proposition 1.} If the following centralized stability condition is satisfied
	\begin{equation}
	\label{E71}
	\begin{cases}
	& \! \! \! \! \! \!
	\|K_C(k)A(k-1)\|_2 \le \lambda <1 \\
	& \! \! \! \! \! \!
	\|K(k)\|_2 \le \eta
	\end{cases} \ \ (\forall k \ge k_0)
	\end{equation}
	where $\eta$ is a finite positive number, then the proposed distributed estimator (\ref{E8}) is stable in the sense of mean-square uniformly bounded (\ref{E15}). \\
		
	Under the distributed communication structure, the knowledge of dynamics and couplings from the overall system is hard to obtain for local estimators. Therefore, the following theorem provides distributed conditions to ensure the stability for all local estimators.\\
	
	\noindent \textbf{Theorem 1 (Distributed stability conditions).} The following distributed conditions are sufficient to ensure the stability for the proposed distributed estimator (\ref{E8}):
	\begin{itemize}
		\item[C1)] For each subsystem
		\begin{equation}
		\label{E25}
		\begin{cases}
		& \! \! \! \! \! \!
		\| K_{C_i}(k) A_i(k-1)\|_2 \le \lambda <1
		\\
		& \! \! \! \! \! \!
		\| K_i(k) \|_2 \le \eta
		\end{cases}
		\ \ (i \in \mathbb{N}_l)
		\end{equation}
		\item[C2)] For each pair of neighbors $(i, i_{\kappa}^{\rho})$
		\begin{equation}
		\label{E26}
		\!\! \epsilon_{i,i_{\kappa}^{\rho}}(k) \epsilon_{i_{\kappa}^{\rho},i}(k) N_{i}(k) \!-\! N_{i,i_{\kappa}^{\rho}}(k) N_{i_{\kappa}^{\rho}}^{-1}(k) N_{i,i_{\kappa}^{\rho}}^{\mathrm{T}}(k) \le 0
		\end{equation}
	\end{itemize}
	where $\eta$ is a finite positive number, while parameter $\epsilon_{i,j}(k)$ satisfies $\sum_{i_{\kappa}^{\rho} \in \Omega_i} \epsilon_{i,i_{\kappa}^{\rho}}(k) =1$ and
	\begin{eqnarray}
	\label{E27}
	\begin{cases}
	& \! \! \! \! \! \!
	N_i(k) : = \begin{bmatrix}
	- \lambda I & K_{C_i}(k) A_i(k-1)\\
	* & - \lambda I
	\end{bmatrix}
	\\
	& \! \! \! \! \! \!
	N_{i,i_{\kappa}^{\rho}}(k) : =
	\begin{bmatrix}
	0 & \!\!\!K_{C_i}(k) A_{i,i_{\kappa}^{\rho}}(k\!-\!1) \\
	A^{\mathrm{T}}_{i_{\kappa}^{\rho},i}(k\!-\!1) K^{\mathrm{T}}_{C_{i_{\kappa}^{\rho}}}(k) & 0
	\end{bmatrix}
	\end{cases}
	\end{eqnarray}
	These conditions are equivalent to the following inequalities:
	\begin{equation}
	\label{E28}
	\begin{cases}
	& \! \! \! \! \! \!
	M_{i,i_{\kappa}^\sigma}(k) \le 0  \ \ ( i \in \mathbb{N}_l, \ i_{\kappa}^{\sigma} \in \Sigma_i )
	\\
	& \! \! \! \! \! \!
	\| K_i(k) \|_2 \le \eta  \ \ (i \in \mathbb{N}_l)
	\end{cases}
	\end{equation}
	where
	\begin{equation}
	\label{E29}
	M_{i,i_{\kappa}^\sigma}(k) \buildrel \Delta \over =
	\begin{bmatrix}
	\epsilon_{i,i_{\kappa}^{\sigma}}(k) N_i(k) & N_{i,i_{\kappa}^\sigma}(k) \\
	* & \epsilon_{i_{\kappa}^{\sigma},i}(k) N_{i_{\kappa}^\sigma}(k)
	\end{bmatrix}
	\end{equation}
	
	\noindent \textbf{Proof.} According to Schur complement lemma [\cite{c36}], the condition (C2) and the first inequality in the condition (C1) are equivalent to $M_{i,i_{\kappa}^\rho}(k) \le 0$ or $M_{i_{\kappa}^\rho,i}(k) \le 0$ for each pair of neighbors $(i,i_{\kappa}^\rho)$. By the definition of $\Sigma_i$ in (\ref{E3}), one can conclude that $M_{i,i_{\kappa}^\sigma}(k) \le 0  \ \ ( i \in \mathbb{N}_l, \ i_{\kappa}^{\sigma} \in \Sigma_i )$.
	Then, define the permutation matrix
	\begin{equation}
	\label{E31}
	Q : = \begin{bmatrix} I_{n_i} & 0 & 0 & 0 \\ 0 & 0 & I_{n_i} & 0 \\ 0 & I_{n_{i_{\kappa}^{\sigma}}} & 0 & 0 \\ 0 & 0 & 0 & I_{n_{i_{\kappa}^{\sigma}}} \end{bmatrix}
	\end{equation}	
	By left and right multiplication of $M_{i,i_{\kappa}^{\sigma}}(k)$ with $Q$ and $Q^{\mathrm{T}}$, the following equivalent inequality is derived:
	\begin{equation}
	\label{E32}
	\begin{aligned}
	\hat M_{i,i_{\kappa}^{\sigma}}(k) \!=\!
	\begin{bmatrix}
	\setlength{\arraycolsep}{2pt}
	U_{i,i_{\kappa}^{\sigma}}(k) &  V_{i,i_{\kappa}^{\sigma}}(k) \\
	* &  U_{i,i_{\kappa}^{\sigma}}(k)
	\end{bmatrix} \le 0 \ \ ( i \in \mathbb{N}_l, i_{\kappa}^{\sigma} \in \Sigma_i )
	\end{aligned}
	\end{equation}
	where
	\begin{eqnarray}
	\label{E33}
	\begin{cases}
	& \! \! \! \! \! \!
	U_{i,i_{\kappa}^{\sigma}}(k) : = 
	\begin{bmatrix}
	- \epsilon_{i,i_{\kappa}^{\sigma}}(k) \lambda I & 0\\
	* & - \epsilon_{i_{\kappa}^{\sigma},i}(k) \lambda I
	\end{bmatrix}
	\\
	& \! \! \! \! \! \!
	\begin{aligned}
	& V_{i,i_{\kappa}^{\sigma}}(k) : = \\
	& \!\! \begin{bmatrix}
	\epsilon_{i,i_{\kappa}^{\sigma}}(k) K_{C_i}(k) A_i(k\!-\!1) & K_{C_i}\!(k) A_{i,i_{\kappa}^{\sigma}}(k\!-\!1) \\
	\!\!\!\! K_{C_{i_{\kappa}^{\sigma}}}\!(k) A_{i_{\kappa}^{\sigma},i}(k\!-\!1) & \!\!\!\! \epsilon_{i_{\kappa}^{\sigma},i}(k) K_{C_{i_{\kappa}^{\sigma}}}(k) A_{i_{\kappa}^{\sigma}}(k\!-\!1)
	\end{bmatrix}
	\end{aligned}
	\end{cases}
	\end{eqnarray}
	By augmenting all the matrices in (\ref{E32}), one has that
	\begin{equation}
	\label{E34}
	\begin{aligned}
	\hat M(k) =\mathrm{diag}\{ & \hat M_{1,1_1^{\sigma}}(k), ..., \hat M_{1,1_{\xi_1}^{\sigma}}(k), \\
	& \hat M_{2,2_1^{\sigma}}(k), ..., \hat  M_{2,2_{\xi_2}^{\sigma}}(k), ...\} \le 0 \\
	\end{aligned}
	\end{equation}
	Then, define the permutation matrix
	\begin{equation}
	\label{E35}
	R : = \mathrm{row}\{e_{1,1_1^{\sigma}}, ..., e_{1,1_{\xi_1}^{\sigma}}, e_{2,2_1^{\sigma}}, ..., e_{2,2_{\xi_2}^{\sigma}}, ...\}
	\end{equation}
	where $e_{i,j} : = \begin{bmatrix} e_i & e_j & 0 & 0 \\ 0 & 0 & e_i & e_j \end{bmatrix}$ and $e_i$ is a matrix with dimension $n \times n_i$ that contains all zero elements, but an identity matrix of dimension $n_i$ at rows $(\sum_{j=1}^{i-1} n_j +1): (\sum_{j=1}^{i} n_j)$. By the property of positive definite matrix, if $\hat M(k)<0$, then the matrix by left and right multiplication of $\hat M(k)$ with $R$ and $R^{\mathrm{T}}$ is negative definite, i.e.,
	\begin{equation}
	\label{E36}
	\begin{aligned}
	& \begin{bmatrix}
	\setlength{\arraycolsep}{2pt}
	\begin{array}{c|c}
	- \lambda I &  \begin{matrix}
	K_{C_1}(k) A_1(k-1) & \cdots & K_{C_1}(k) A_{1,l}(k-1) \\
	K_{C_2}(k) A_{2,1}(k-1) & \cdots & K_{C_2}(k) A_{2,l}(k-1) \\
	\vdots & \ddots & \vdots \\
	K_{C_l}(k) A_{l,1}(k-1) & \cdots & K_{C_l}(k) A_l(k-1) \\ \end{matrix} 
	\\  \hline
	* & - \lambda I
	\end{array}
	\end{bmatrix}\\
	& \le 0 
	\end{aligned}
	\end{equation}
	Inequality (\ref{E36}) is equivalent to
	\begin{equation}
	\label{E37}
	\begin{bmatrix}
	- \lambda I & K_C(k) A(k-1)  \\
	* & - \lambda I    
	\end{bmatrix}\le 0
	\end{equation}
	By Schur complement lemma, one has that
	\begin{equation}
	\label{E38}
	\| K_C(k)A(k-1) \|_2 \le \lambda
	\end{equation}
	According to Proposition 1, inequality (\ref{E38}) and the second inequality in the condition (C1) are sufficient to ensure the mean-square boundedness (\ref{E15}). This completes the proof. \\
	
	\noindent \textbf{Remark 4.} Intuitively, the stability of an independent subsystem without any couplings is not influenced by its neighboring subsystems, and thus local mean-square uniform boundedness condition is enough to ensure the stability. However, the stability condition for the subsystem that coupled with its neighbors will be tighter than the local mean-square uniform boundedness condition. Therefore, the distributed stability conditions in Theorem 1 contain two parts, the condition (C1) ensures that local estimation error system without interconnected terms is stable, while the condition (C2) is an additional requirement for the stability of systems with some coupling relationships. \\
	
	\noindent \textbf{Remark 5.} For the distributed control problem of interconnected systems with known system states, the distributed conditions for stabilizing systems can be obtained by a similar derivation of Theorem 1. When the process dynamics (\ref{E1}) is controlled by an additional input term ``$B_i(k) u_i(k)$", the distributed state feedback controllers $u_i(k)=-\sum_{i_{\kappa}^{\rho} \in \Omega_i} K^u_{i,i_{\kappa}^{\rho}}(k) x_i(k)$ and ``$u_i(k)=-K^u_i(k) x_i(k)$" can be designed, then the distributed stabilization conditions can be obtained by decomposing the matrix inequality $\|A(k)-B(k) K^u(k)\|_2 \le \lambda$ with the property of positive definite matrix, where $B(k)$ and $K^u(k)$ are the corresponding augmented matrices.\\
	
	The determination of the parameter $\lambda$ is a trade-off between the stability and the performance of estimators, where smaller $\lambda$ can provide more conservative margin of distributed stability and potentially worse estimation performance. On the other hand, the parameter $\eta$ only influences the ultimate boundedness of estimators and can be chosen as a large number to avoid estimation performance degradation.
	
	Notice that the above distributed stability conditions need subsystem $\mathbf{S}_i$ to know the gain $K_{i_{\kappa}^{\rho}}(k)$ from subsystem $\mathbf{S}_{i_{\kappa}^{\rho}}$ timely. However, one-step communication delay, naturally risen from networked environments, is inevitable and need to be taken into account when the estimator gain information is transmitted over the communication network from neighboring subsystems. To extend the result of Theorem 1 to more general communication environments with one-step transmission delay, the following conditions without synchronously knowing neighboring estimator gains are further proposed.\\
	
	\noindent \textbf{Corollary 1.} For each subsystem, if the following inequalities are satisfied:
	\begin{equation}
	\label{E39}
	\begin{cases}
	& \! \! \! \! \! \!
	\|K_{C_i}(k)\|_2 \le \beta_i
	\\
	& \! \! \! \! \! \!
	\| K_i(k) \|_2 \le \eta
	\end{cases}
	\ \ (i \in \mathbb{N}_l)
	\end{equation}
	where $\eta$ is a finite positive number, and $\beta_i \le \frac{\lambda}{\alpha_i}$ is constrained by
	\begin{equation}
	\label{E40}
	\left(\alpha_i \beta_i - \lambda \right) \left(\alpha_{i_{\kappa}^{\rho}} \beta_{i_{\kappa}^{\rho}} - \lambda \right) \ge \frac{\beta_i^2 \alpha^2_{i,i_{\kappa}^{\rho}}}{\bar \epsilon_{i,i_{\kappa}^{\rho}} \bar \epsilon_{i_{\kappa}^{\rho},i}} \ \ i_{\kappa}^{\rho} \in \Omega_i
	\end{equation}
	with parameter $\bar \epsilon_{i,j}$ satisfies $\sum_{i_{\kappa}^{\rho} \in \Omega_i} \bar \epsilon_{i,i_{\kappa}^{\rho}} =1$, then the proposed distributed estimator is stable in the sense of mean-square uniformly bounded (\ref{E15}). \\
	
	\noindent \textbf{Proof.} 
	The following upper bounds can be derived from the inequality (\ref{E39}):
	\begin{equation}
	\label{E41}
	\begin{cases}
	& \! \! \! \! \! \!
	\|K_{C_i}(k) A_i(k-1)\|_2 \le \beta_i \alpha_i \le \lambda
	\\
	& \! \! \! \! \! \!
	\|K_{C_i}(k) A_{i,i_{\kappa}^{\rho}}(k-1)\|_2 \le \beta_i \alpha_{i,i_{\kappa}^{\rho}}
	\\
	& \! \! \! \! \! \!
	\|K_{C_{i_{\kappa}^{\rho}}}(k) A_{i_{\kappa}^{\rho},i}(k-1)\|_2 \le \beta_i \alpha_{i_{\kappa}^{\rho},i}
	\end{cases}
	\end{equation}
	By the Schur complement lemma, the first inequality in (\ref{E41}) can be converted to
	\begin{equation}
	\label{E42}
	\begin{bmatrix}
	- \beta_i \alpha_i I & K_{C_i}(k) A_i(k-1)\\
	* & - \beta_i \alpha_i I
	\end{bmatrix} \le 0
	\end{equation}
	Hence, the upper bounds of $N_i(k)$ and $N_i^{-1}(k)$ can be obtained as
	\begin{equation}
	\label{E43}
	\begin{cases}
	& \! \! \! \! \! \!
	N_i(k) \le  \left( \beta_i \alpha_i - \lambda \right) I
	\\
	& \! \! \! \! \! \!
	N^{-1}_i(k) \ge \frac{1}{\beta_i \alpha_i - \lambda} I
	\end{cases}
	\end{equation}
	By the second and the third inequalities of (\ref{E41}), one has the following inequality:
	\begin{equation}
	\label{E44}
	N_{i,i_{\kappa}^{\rho}}(k) N_{i,i_{\kappa}^{\rho}}^{\mathrm{T}}(k) \le
	\begin{bmatrix}
	\beta_i^2 \alpha^2_{i,i_{\kappa}^{\rho}} I & 0 \\
	0 & \beta_{i_{\kappa}^{\rho}}^2 \alpha^2_{i_{\kappa}^{\rho},i} I
	\end{bmatrix}
	\end{equation}
	Therefore, it can be concluded that
	\begin{equation}
	\label{E45}
	\begin{aligned}
	& \bar \epsilon_{i,i_{\kappa}^{\rho}}  \bar \epsilon_{i_{\kappa}^{\rho},i} N_{i}(k) - N_{i,i_{\kappa}^{\rho}}(k) N_{i_{\kappa}^{\rho}}^{-1}(k) N_{i,i_{\kappa}^{\rho}}^{\mathrm{T}}(k) \\
	& \le \bar \epsilon_{i,i_{\kappa}^{\rho}}  \bar \epsilon_{i_{\kappa}^{\rho},i} \left( \beta_i \alpha_i - \lambda \right) I  \\
	& \ \ \ \ -  \frac{1}
	{\alpha_{i_{\kappa}^{\rho}} \beta_{i_{\kappa}^{\rho}} - \lambda} \begin{bmatrix}
	\beta_i^2 \alpha^2_{i,i_{\kappa}^{\rho}} I & 0 \\
	0 & \beta_{i_{\kappa}^{\rho}}^2 \alpha^2_{i_{\kappa}^{\rho},i} I
	\end{bmatrix}
	\end{aligned}
	\end{equation}
	Under the constraints in (\ref{E40}) and taking $\epsilon_{i,i_{\kappa}^{\rho}}(k) = \bar \epsilon_{i,i_{\kappa}^{\rho}}$, the condition (C2) in Theorem 1 is derived. This completes the proof. \\
	
	To balance the stability margin of each subsystem, the parameter $\epsilon_{i,j}(k)$ and $\bar \epsilon_{i,j}$ should be proportional to the size of couplings, and a feasible parameter selection is given by
	\begin{equation}
	\begin{cases}
	& \! \! \! \! \! \!
	\epsilon_{i,j}(k) = \frac{\|A_{i,j}(k-1)\|_2 + \|A_{j,i}(k-1)\|_2}{\sum_{i_{\kappa}^{\rho} \in \Omega_i} \left( \|A_{i,i_{\kappa}^{\rho}}(k-1)\|_2 + \|A_{i_{\kappa}^{\rho},i}(k-1)\|_2 \right) }
	\\
	& \! \! \! \! \! \!
	\bar \epsilon_{i,j} = \frac{\alpha_{i,j} + \alpha_{j,i}}{\sum_{i_{\kappa}^{\rho} \in \Omega_i} \left( \alpha_{i,i_{\kappa}^{\rho}} + \alpha_{i_{\kappa}^{\rho},i} \right) }
	\end{cases}
	\end{equation}
		
	\noindent \textbf{Remark 6.} By Corollary 1, the distributed stability conditions are simplified into finding an appropriate time-invariant parameter $\beta_i$. Unlike the conditions in [\cite{c33}] that require huge communication burden to exchange message matrices among subsystems, the calculation of $\beta_i$ in this paper only needs subsystems to communicate with their neighbors to exchange the knowledge of $\beta_{i_{\kappa}^{\rho}}$ and $\alpha_{i_{\kappa}^{\rho}}$. Therefore, this procedure can be achieved offline with less communication overhead. Notice that the stability result with less communication and computational burden is more suitable for time-varying interconnected systems with different couplings and dynamics at each instant. \\
	
	The distributed calculation of $\beta_i$ can be implemented in the following Algorithm.
	
	\begin{algorithm}[H]
		\caption{Distributed calculation for $\beta_i$}
		\label{Algo:1}
		\begin{algorithmic}[1]
			\FOR{$i := 1$ \TO $L$}{
				\IF{$i \neq 1$}
				\STATE Subsystem $\mathbf{S}_i$ receives $\beta_{i_{\kappa}^{\rho}}$ and $\alpha_{i_{\kappa}^{\rho}}$ from subsystem $\mathbf{S}_{i_{\kappa}^{\rho}} \ (i_{\kappa}^{\rho} \in \Omega_i \cap \mathbb{N}_{i-1})$;
				\ENDIF
				\STATE Subsystem $\mathbf{S}_i$ calculates $\beta_i<\lambda$ that satisfies (\ref{E40}) for each coupling pair $(i,i_{\kappa}^{\rho}), i_{\kappa}^{\rho} \in \Omega_i \cap \mathbb{N}_{i-1}$;
				\STATE Subsystem $\mathbf{S}_i$ sends the calculated $\beta_i$ and $\alpha_i$ to subsystem $\mathbf{S}_{i_{\kappa}^{\sigma}} \ i_{\kappa}^{\sigma} \in  \Sigma_i$;
			}
			\ENDFOR
		\end{algorithmic}
	\end{algorithm}
	
	\noindent \textbf{Remark 7.} The small gain theorem for interconnected systems can be stated as follows. Suppose that each local estimation error system (\ref{E9}) satisfies the local mean-square uniform boundedness condition $\|K_{C_i}(k) A_i (k-1)\|_2<1$, then the augmented estimation error system in (\ref{E10}) is stable if the set of small gain conditions $\|A_{i_1,i_2} A_{i_2,i_3} ... A_{i_r,i_1}\|<1$ ($1\le i_s \le l$, $i_s \neq i_{s'}$ if $s \neq s'$) holds for each $r=2,...,l$. The small gain conditions mean that the composition of the coupling matrices along every closed cycle is stable. However, it is hard to apply the small gain theorem to design distributed estimator or controller for interconnected systems with arbitrary coupling structures. For distributed estimation problem, feedback is introduced to adjust the size of $K_{C_i}(k) A_i (k-1) \ (i \in \mathbb{N}_l)$ in a distributed manner such that the augmented estimation error system is stable, while the coupling matrices $A_{i,i_{\kappa}^\rho}(k)$ cannot be adjusted. The small gain theorem requires that $\|A_{i_1,i_2} A_{i_2,i_3} ... A_{i_r,i_1}\|<1$, which is not always satisfied and irrelevant to the estimator design. A natural problem is what distributed conditions does a subsystem need to meet with its neighbors such that the overall system is stable. To address the above problem, the distributed stability conditions are derived by decomposing the centralized stability condition $\|K_C(k)A(k-1)\|_2<\lambda$ in the paper. The result in Theorem 1 turns out to be the matrix inequalities for each pair of neighbors. Therefore, each subsystem only needs to satisfy these matrix inequalities with its neighbors, then the stability for the overall system can be ensured.\\
	
	\noindent \textbf{Remark 8.} Compared with the stability analysis by Lyapunov functions [\cite{c1, c27, c28}], the proposed distributed stability conditions in Corollary 1 are less conservative in the requirement of weak coupling assumptions. According to the inequality (\ref{E40}), the strength of coupling $\alpha_{i,i_{\kappa}^{\rho}}$ can be arbitrarily large as long as the stability parameter $\beta_i$ is designed small enough. On the other hand, the conditions in Corollary 1 can be directly applied to the distributed estimation problem when the parameters are determined by Algorithm 1 offline.
	
	\subsection{Optimization-based Distributed Estimator}
	In what follows, we would like to design optimal estimators for time-varying interconnected systems in a distributed way. We have the following results on optimization-based distributed estimator design. First of all, let us define the following matrices:
	\begin{eqnarray}
	\label{E46}
	\begin{cases}
	& \! \! \! \! \! \! \!
	\mathcal{D}_{P_i}(k) : = \mathrm{col} \left\{ \sqrt{\left[P_i(k)\right]_1}, ..., \sqrt{\left[P_i(k)\right]_{n_i}} \right\}
	\\
	& \! \! \! \! \! \! \!
	\mathcal{D}_{\hat P_i}(k) : = \mathrm{col} \left\{ \sqrt{\left[\hat P_i(k)\right]_1}, ..., \sqrt{\left[\hat P_i(k)\right]_{n_i}} \right\}
	\end{cases}
	\end{eqnarray}
	where $\hat P_i(k)$ is an upper bound of $P_i(k)$ and $\left[P_i(k)\right]_{\tau}$ is the $\tau$th diagonal element of $P_i(k)$. Then, the gain design for the proposed distributed estimator (\ref{E8}) is provided in the following Theorem. \\
	
	\noindent \textbf{Theorem 2.} For the time-varying interconnected system (\ref{E1}), the gain matrix $K_i^{\mathrm{opt}}(k)$ of the proposed distributed estimator (\ref{E8}) is obtained by minimizing an upper bound of estimation error covariance and keeping the designed estimator mean-square uniformly bounded, as the following optimization problem:
	\begin{equation}
	\label{E47}
	\begin{aligned}
	&\min_{K_i(k)} \mathrm{Tr}\{\hat G_i(k)\} \\
	&\mathrm{s.t.} \
	\begin{cases}
	& \! \! \! \! \! \!
	\begin{bmatrix}
	& \!\!\!\!\!\! - \hat G_i(k) & K_{C_i}(k) \hat P^p_i(k) & K_i(k) D_i(k) Q_{v_i} \\
	& * & - \hat P^p_i(k)& 0 \\
	& * & * & - Q_{v_i}  \\
	\end{bmatrix} \!\!<\!0
	\\
	& \! \! \! \! \! \!
	(\ref{E28}) \ \ \text{or} \ \ (\ref{E39})
	\end{cases}
	\end{aligned}
	\end{equation}
	where $\hat P^p_i(k)$ is an upper bound of one-step prediction error covariance and is calculated as
	\begin{eqnarray}
	\label{E48}
	\begin{aligned}
	& \hat P^p_i (k) = A_i(k-1) \hat P_i(k-1) A_i^{\mathrm{T}}(k-1)  \\
	& + \! \sum_{i_{\kappa}^{\rho} \in \Omega_i} \! A_i(k\!-\!1) \mathcal{D}_{\hat P_i}(k-1) \mathcal{D}^{\mathrm{T}}_{\hat P_{i_{\kappa}^{\rho}}}(k-1) A_{i,i_{\kappa}^{\rho}}^{\mathrm{T}}(k-1) \\
	& + \! \sum_{i_{\kappa}^{\rho} \in \Omega_i} \! A_{i,i_{\kappa}^{\rho}}(k\!-\!1) \mathcal{D}_{\hat P_{i_{\kappa}^{\rho}}}(k-1) \mathcal{D}^{\mathrm{T}}_{\hat P_i}(k-1) A_i^{\mathrm{T}}(k\!-\!1)  \\
	& + \!\! \sum_{i_{\kappa_1}^{\rho} \in \Omega_i} \sum_{i_{\kappa_2}^{\rho} \in \Omega_i} \left\{ A_{i,i_{\kappa_1}^{\rho}}(k-1) \mathcal{D}_{\hat P_{i_{\kappa_1}^{\rho}}}(k-1) \right. \\
	& \left.  \ \ \ \ \ \ \ \ \ \ \ \ \ \ \ \ \ \ \ \  \times \mathcal{D}^{\mathrm{T}}_{\hat P_{i_{\kappa_2}^{\rho}}}(k-1) A_{i,i_{\kappa_2}^{\rho}}^{\mathrm{T}}(k-1) \right\} \\
	& + \Gamma_i(k-1) Q_{w_i} \Gamma_i^{\mathrm{T}}(k-1)
	\end{aligned}
	\end{eqnarray}
	with the upper bound of estimation error covariance $\hat P_i(k-1)$ calculated as
	\begin{eqnarray}
	\label{E49}
	\begin{aligned}
	& \hat P_i(k-1) = \left[I-K_i^{\mathrm{opt}}(k-1) C_i(k-1) \right] \hat P_i^p(k-2) \\
	& \times \left[I-K_i^{\mathrm{opt}}(k-1) C_i(k-1) \right]^{\mathrm{T}} \\
	& +  K_i^{\mathrm{opt}}(k\!-\!1) D_i(k\!-\!1) Q_{v_i} D_i^{\mathrm{T}}(k\!-\!1) \!\! \left[K_i^{\mathrm{opt}}(k\!-\!1)\right]^{\mathrm{T}}
	\end{aligned}
	\end{eqnarray}
	
	\noindent \textbf{Proof.} 
	The cross-covariances $P_{i,j}(k)$ among subsystems are difficult to online calculate by local communication, which means that direct calculation of the one-step prediction error covariance $P^p_i(k)$ in (\ref{E50}) is not feasible. Therefore, an upper bound of the estimation error covariance $\hat P_i(k) \ge P_i(k)$ is constructed and used for the gain design problem. Let $\left[ \tilde x_i(k) \right]_{\tau_1} \in \mathbb{R}$ be the $\tau_1$th component of $\tilde x_i(k)$, while $\left[ \tilde x_j(k) \right]_{\tau_2}$ is defined as the $\tau_2$th component of $\tilde x_j(k)$. By resorting to the well-known H\"{o}lder inequality, one has that
	\begin{equation}
	\label{E51}
	\begin{aligned}
	\mathrm{E}\left\{\{ \left[ \tilde x_i(k) \right]_{\tau_1} \left[ \tilde x_j(k) \right]_{\tau_2} \right\}  	
	& \le \mathrm{E}\left\{ \left\lvert \left[ \tilde x_i(k) \right]_{\tau_1} \left[ \tilde x_j(k) \right]_{\tau_2} \right\rvert \right\} \\
    & \! \! \! \! \! \! \! \! \! \! \! \! \! \le \sqrt{\mathrm{E}\left\{ \left[ \tilde x_i(k) \right]_{\tau_1}^2 \right\}} \sqrt{\mathrm{E}\left\{ \left[ \tilde x_j(k) \right]_{\tau_2}^2 \right\}}
	\end{aligned}
	\end{equation}
	Thus, the following upper bound of $P_{i,j}(k)$ is derived:
	\begin{equation}
	\label{E52}
	P_{i,j}(k) \le \mathcal{D}_{P_i}(k) \mathcal{D}^{\mathrm{T}}_{P_j}(k)
	\end{equation}
    Then, applying the inequality (\ref{E52}) to (\ref{E50}), it turns out that
	\begin{eqnarray}
	\label{E53}
	\begin{aligned}
	& P_i^p(k) \le A_i(k-1) P_i(k-1) A_i^{\mathrm{T}}(k-1)  \\
	& + \! \sum_{i_{\kappa}^{\rho} \in \Omega_i} \! A_i(k\!-\!1) \mathcal{D}_{P_i}(k-1) \mathcal{D}^{\mathrm{T}}_{P_{i_{\kappa}^{\rho}}}(k-1)  A_{i,i_{\kappa}^{\rho}}^{\mathrm{T}}(k-1) \\
	& + \! \sum_{i_{\kappa}^{\rho} \in \Omega_i} \!\! A_{i,i_{\kappa}^{\rho}}(k\!-\!1) \mathcal{D}_{P_{i_{\kappa}^{\rho}}}(k-1) \mathcal{D}^{\mathrm{T}}_{P_i}(k-1)  A_i^{\mathrm{T}}(k\!-\!1)  \\
	& + \! \sum_{i_{\kappa_1}^{\rho} \in \Omega_i} \sum_{i_{\kappa_2}^{\rho} \in \Omega_i} \left\{ A_{i,i_{\kappa_1}^{\rho}}(k-1) \mathcal{D}_{P_{i_{\kappa_1}^{\rho}}}(k-1) \right. \\
	& \left.  \ \ \ \ \ \ \ \ \ \ \ \ \ \ \ \ \ \ \ \    
	\times \mathcal{D}^{\mathrm{T}}_{P_{i_{\kappa_2}^{\rho}}}(k-1) A_{i,i_{\kappa_2}^{\rho}}^{\mathrm{T}}(k-1) \right\} \\
	& + \Gamma_i(k-1) Q_{w_i} \Gamma_i^{\mathrm{T}}(k-1)  \\
	\end{aligned}
	\end{eqnarray}
	Therefore, an upper bound of $P_i^p(k)$ is constructed by $P_i^p(k) \le \hat P_i^p(k)$. In this case, an upper bound of local estimation error covariance is derived as
	\begin{equation}
	\label{E55}
	\begin{aligned}
	\hat P_i(k) = 
	& K_{C_i}(k) \hat P_i^p(k) \left[ K_{C_i}(k) \right]^{\mathrm{T}}  \\
	& +  K_i(k) D_i(k) Q_{v_i} D_i^{\mathrm{T}}(k) K_i^{\mathrm{T}}(k)
	\end{aligned} 
	\end{equation}
	Then, it is proposed to construct an upper bound of $\hat P_i(k)$ as $\hat G_i(k)$ satisfying 
	\begin{equation}
	\label{E56}
	\hat P_i(k) - \hat G_i(k) <0
	\end{equation}
	The optimal estimator gain is obtained by minimizing this upper bound $\hat G_i(k)$, which turns to be an optimization problem:
	\begin{equation}
	\label{E57}
	\begin{aligned}
	& K_i^{\mathrm{opt}}(k) = \arg \min_{K_i(k)} \mathrm{Tr}\{\hat G_i(k)\} \\
	& \mathrm{s.t.} \ \hat P_i(k) - \hat G_i(k) <0
	\end{aligned}
	\end{equation}
	By Schur complement lemma, the inequality constraint in (\ref{E57}) is converted into
	{\small{
			\begin{eqnarray}
			\label{E58}
			\begin{bmatrix}
			K_i(k) D_i(k) Q_{v_i} D_i^{\mathrm{T}}(k) K_i^{\mathrm{T}}(k) - G_i(k) & K_{C_i}(k) \\
			* & -\left[ \hat P^p_i(k) \right]^{-1}
			\end{bmatrix}<0
			\end{eqnarray}}}Then, 
	the first inequality constraint in (\ref{E47}) is further derived by using Schur complement lemma again. Adding the distributed stability constraints in Theorem 1 or Corollary 1, the optimization problem in Theorem 2 is formulated. This completes the proof. \\
	
	\noindent \textbf{Remark 9.} The inequality constraints (\ref{E28}) and (\ref{E39}) can be rewritten as linear matrix inequality forms, so the optimization problem in Theorems 2 can be directly solved by the function ``mincx" of MATLAB LMI toolbox [\cite{c36}]. In addition, the information used in the optimization problem (\ref{E47}) is from subsystem $\mathbf{S}_i$ and its neighbors, and the computational complexity is mainly determined by the dimensions of these subsystems. Therefore, the proposed estimators are recursive and fully distributed such that can be deployed for large-scale interconnected systems with local communication and computation requirements.\\
	
	\noindent \textbf{Remark 10.} Notice that the work in [\cite{c19, c20}] only provides stability conditions with specific subsystem coupling structures, while the designed distributed estimators in Theorem 2 directly use the newly proposed subsystem-level stability conditions for general interconnected systems. This design methdology that combines the optimality and stability can overcome the disadvantages of totally local estimator analysis in terms of the stability problem. Moreover, the developed stability conditions enable plug-and-play operations, which means newly added subsystem does not influence the stability of previous subsystems and its own stability can be ensured by collecting its neighbors' information. Thus, there is no need to redesign the stability parameters and this property is helpful for deployment of distributed estimators. \\
	
	\begin{algorithm}[]
		\caption{Distributed Estimation for Time-varying interconnected systems}
		\label{Algo:2}
		\begin{algorithmic}[1]
			\IF{Communication is one-step delayed}{
				\STATE Offline calculation of $\beta_i$ by Algorithm 1;
			}
			\ENDIF
			\FOR{$i := 1$ \TO $L$}{
				\STATE Subsystem $\mathbf{S}_i$ collects local measurement $y_i(k)$, neighbors' estimated states $\hat x_{i_{\kappa}^\rho}(k-1)$ and error covariance bounds $\hat P_{i_{\kappa}^{\rho}}(k\!-\!1)$, and $K_{i_{\kappa}^\sigma}(k) \ \ (i_{\kappa}^\sigma \in \Sigma_i (k-1))$;
				\STATE Calculate $\hat P^p_i (k)$ by (\ref{E48});
				\IF{Communication is one-step delayed}{
					\STATE Determine the estimator gain $K_i(k)$ by solving the optimization problem (\ref{E47}) with constraints in (\ref{E39});
				}
				\ELSE
					\STATE Determine the estimator gain $K_i(k)$ by solving the optimization problem (\ref{E47}) with constraints in (\ref{E28});
				\ENDIF
				\STATE Calculate $\hat P_i(k)$ by (\ref{E49});
				\STATE Calculate the distributed estimate $\hat x_i(k)$ by (\ref{E8});
				\STATE Subsystem $\mathbf{S}_i$ sends the calculated $\hat x_i(k)$, $\hat P_i(k)$ (only for Gaussian noise situation) and $K_i(k)$ to its neighbors. 
			}
			\ENDFOR
			\STATE Return to Step 4 and implement Steps 4-15 for calculating $\hat x_i(k+1) (i=1,2,...,l)$.
		\end{algorithmic}
	\end{algorithm}

	From Theorem 2, the computational procedures of the distributed estimation for general interconnected systems with and without one-step communication delay can be summarized by Algorithm 2. For systems with ideal communication, the real-time transmission of estimator gains is feasible and the required information for the inequality constraints in (\ref{E28}) can be obtained timely. However, the stability conditions in (\ref{E28}) will not work any more when one-step communication delay is taken into consideration. Instead, offline calculation of $\beta_i$ for the inequality constraints in (\ref{E39}) can solve the problem caused by communication delay.
	
	\begin{figure}[]
		\begin{center}
			\includegraphics[height=3cm, width=8cm]{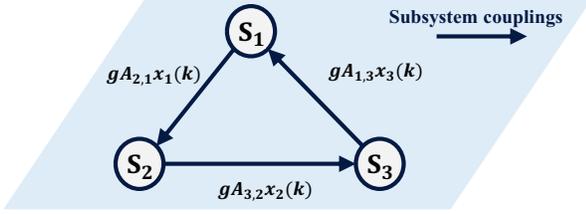}    
			\caption{The couplings among subsystems for an interconnected system.}  
			\label{fig2}                                 
		\end{center}                                 
	\end{figure}
	\section{Simulation Examples}
	\begin{figure*}[]
		\begin{center}
			\includegraphics[height=6cm, width=14cm]{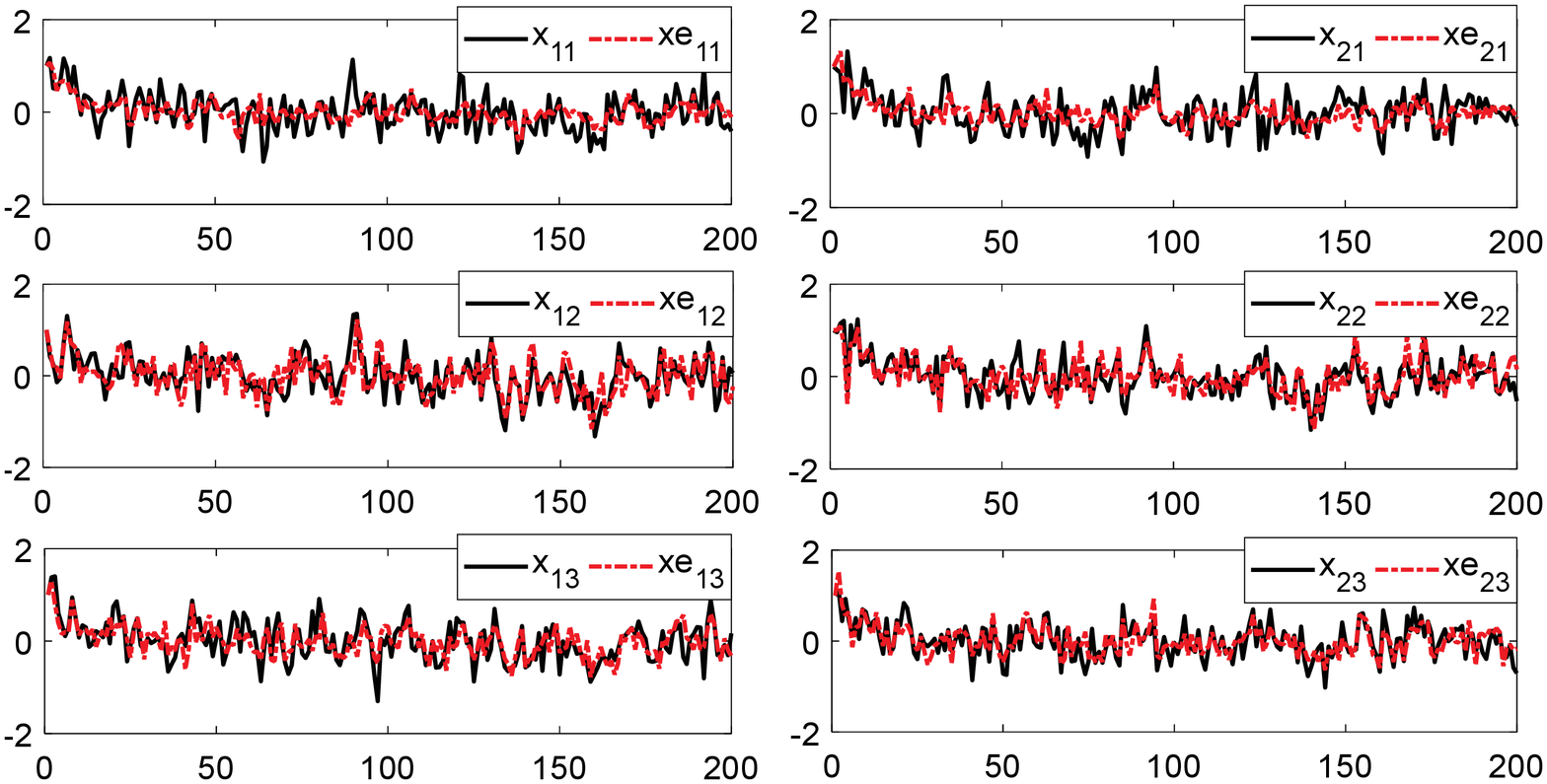}    
			\caption{The trajectories of the states and the corresponding estimated values by Theorem 2 for the interconnected system.}  
			\label{fig3}                                 
		\end{center}                                 
	\end{figure*}
	\begin{figure}[]
		\begin{center}
			\includegraphics[height=2.2cm, width=9cm]{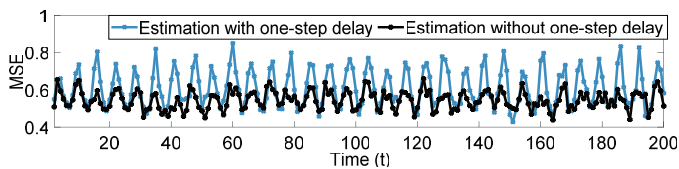}    
			\caption{MSE comparison of the distributed estimators in Algorithms 2 with and without communication delay.}  
			\label{fig4}                                 
		\end{center}                                 
	\end{figure}
	\begin{figure}[]
		\begin{center}
			\includegraphics[height=2.2cm, width=9cm]{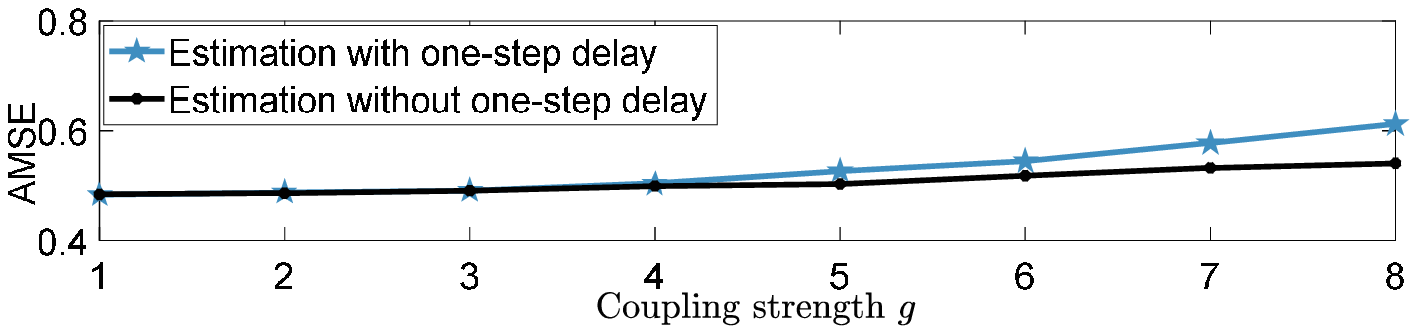}
			\caption{AMSE performance under different coupling strengths.}
			\label{fig5}   
		\end{center}                                 
	\end{figure}
	To illustrate the effectiveness of the proposed distributed estimators, a numerical study result is reported in this section. Let us consider the following interconnected system with three subsystems:
	\begin{equation}
	\label{E66}
	\mathbf{S}\!:\!
	\begin{cases}
	& \! \! \! \! \! \!
	x_1(k+1) \!=\! A_1(k) x_1(k) + gA_{1,3} x_3(k) + \Gamma_1 w_1(k)
	\\
	& \! \! \! \! \! \!
	x_2(k+1) \!=\! A_2(k) x_2(k) + gA_{2,1} x_1(k) + \Gamma_2 w_2(k)
	\\
	& \! \! \! \! \! \!
	x_3(k+1) \!=\! A_3(k) x_3(k) + gA_{3,2} x_2(k) + \Gamma_3 w_3(k)
	\\
	& \! \! \! \! \! \!
	y_1(k) = C_1(k) x_1(k) + D_1 v_1(k)
	\\
	& \! \! \! \! \! \!
	y_2(k) = C_2(k) x_2(k) + D_2 v_2(k)
	\\
	& \! \! \! \! \! \!
	y_3(k) = C_3(k) x_3(k) + D_3 v_3(k)
	\end{cases}
	\end{equation}
	where
	\begin{equation}
	\label{E67}
	\begin{cases}
	& \! \! \! \! \! \!
	A_1(k) = \begin{bmatrix} 0.2 & 0.2+0.2 \cos(k) \\ 0.2+0.1 \sin(k) & 0.2 \end{bmatrix}
	\\
	& \! \! \! \! \! \!
	A_2(k) = \begin{bmatrix} 0.3 & 0.1+0.3 \cos(k) \\ 0.2+0.2 \sin(k) & 0.2 \end{bmatrix}
	\\
	& \! \! \! \! \! \!
	A_3(k) = \begin{bmatrix} 0.3 & 0.1+0.2 \sin(k) \\ 0.1+0.1 \cos(k) & 0.2 \end{bmatrix}
	\\
	& \! \! \! \! \! \!
	C_1(k) = \begin{bmatrix} 0.3+0.3 \cos(k) & 0.4 \end{bmatrix}
	\\
	& \! \! \! \! \! \!
	C_2(k) = \begin{bmatrix} 0.6+0.2 \cos(k) & 0.3 \\ 0.2 & 0.7+0.1 \sin(k) \end{bmatrix}
	\\
	& \! \! \! \! \! \!
	C_3(k) = \begin{bmatrix} 0.5+0.1 \sin(k) & 0.3 \\ 0.1 & 0.7+0.1 \cos(k) \end{bmatrix}
	\\
	& \! \! \! \! \! \!
	A_{1,3} = A_{2,1} = A_{3,2} = \begin{bmatrix}0.1 & 0 \\ 0 & 0.1\end{bmatrix}
	\end{cases}
	\end{equation}
	and $\Gamma_i$ and $D_i$ are identity matrices. The parameter $g$ is used to adjust the strength of couplings, and the coupling structure is described in Fig. \ref{fig2}. The process noise $w_i(k) \ (i \in \{1,2,3\})$ is Gaussian noises with covariance $\mathrm{diag}\{0.1,0.1\}$, and the measurement noises $v_i(k) \ (i=1,2,3)$ are Gaussian noises with covariances $0.1$, $\mathrm{diag}\{0.1,0.1\}$ and $\mathrm{diag}\{0.1,0.1\}$, respectively. The values for $\beta_i$ are calculated for different coupling strengths by tuning the parameter $g$, and the result is shown in Table 1. 
	\begin{table}[]
		\renewcommand{\arraystretch}{1.3}
		\caption{Values of $\beta_i$ under Different Coupling Strength}  
		\centering
		\label{table1}
		\begin{tabular}{|c|c|c|c|c|c|c|c|c|c|}
			\hline 
			\thead[l]{Coupling\\Strength $g$} & 0.5 & 1 & 1.5 & 2 & 2.5 & 3 & 3.5 & 4 \\
			\hline
			$\beta_1$ & 1.08 & 1.08 & 1.08 & 1.08 & 1.08 & 1.08 & 1.08 & 1.08\\[-1pt]  
			\hline 
			$\beta_2$ & 1.21 & 1.01 & 0.90 & 0.81 & 0.75 & 0.70 & 0.66 & 0.63\\[-1pt]
			\hline 
			$\beta_3$ & 1.84 & 1.57 & 1.36 & 1.19 & 1.05 & 0.94 & 0.86 & 0.78\\[-1pt]
			\hline
		\end{tabular}
	\end{table}
	As the connection of subsystems gets more and more strong, the calculated value of $\beta_i$ decreases such that the stability conditions become stricter. This observation is consistent with the fact that the convergence rate and stability margin of the overall system is related to the size of its transition matrix.
	
	Then, the proposed optimization-based distributed estimators are deployed to estimate the states of this interconnected system. Under one-step communication delay, the trajectories of the states and the corresponding estimated values by Theorem 2 for this interconnected system are plotted in Fig. \ref{fig3} when $g=4$. As shown in Fig. \ref{fig3}, the proposed distributed estimator can track the real states well under a large coupling strength with communication delay. To compare the performance of the distributed estimators with and without the influence of one-step communication delay, Monte Carlo simulations with 100 runs have been performed by randomly varying the realization of process and measurement noises. The mean square error (MSE) is introduced to evaluate the performance, where
	\begin{equation}
	\label{E68}
	\mathrm{MSE}(k) = \sum_{s=1}^S \frac{\|e_s(k)\|^2}{S}
	\end{equation}
	with $e_s(t)$ being the state estimation error at the instant $k$ in the $s$th simulation. Fig. \ref{fig4} depicts the MSE performance comparison for two cases in Algorithms 2. The result shows that the estimation accuracy can maintain at a satisfactory level when one-step communication delay is taken into consideration. Moreover, to evaluate the dependence of performance on different coupling strengths, the AMSE (i.e., the asymptotic MSE defined as the average of the MSE computed in the whole time interval) is reported in Fig. \ref{fig5}. As the stability constraints are getting stricter under stronger couplings, the estimation accuracy is becoming worse. The performance degradation is caused by its conservatism from time-invariant stability parameters.
	
	\section{Conclusion}
	In this paper, we presented new results for subsystem-level stability analysis and distributed estimator design for time-varying interconnected systems with arbitrary coupling structures. The proposed distributed stability conditions can ensure mean-square uniform boundedness without the requirement for the knowledge of dynamics and couplings from the overall interconnected systems. Then, the simplified conditions that do not need real-time exchange of subsystems' gain information were developed for systems with one-step communication delay. Particularly, we showed that the distributed stability conditions do not need any coupling structure assumption and can be easily extended when a new subsystem is added to the original interconnected system. These conditions are applied to distributed estimator design problem for time-varying interconnected systems, and novel optimization-based estimator design approaches were proposed. Notice that the designed estimators are fully distributed, where only local information and the information from neighbors are required for the estimator iteration form and the stability conditions. Finally, an illustrative example was employed to show the effectiveness of the proposed methods. 
    
    Several topic for future research is left open. Extensions of the presented distributed stability conditions to co-design of distributed estimator and controller will be important. Another interesting extension is the development of secure estimator for interconnected systems with stability constraints. Due to the frequent information exchange among subsystems and the broadcast nature of communication medium, it is more vulnerable for practical interconnected systems to various attacks. To prevent system information from being collected by eavesdroppers to generate sophisticated attacks, the design of defense mechanisms is required and will be one of our future work. Meanwhile, the influence of cyber attacks will propagate among subsystems, and how to detect cyber attacks by subsystem cooperation is an important and interesting problem.

    \section{Appendix}
    \noindent \textbf{Proof of Proposition 1.} From the conditions in (\ref{E71}), an upper bound of $K_C(k)$ can be derived as
    \begin{equation}
    \label{E19}
    \| K_C(k) \|_2 < 1+ \eta \delta_c,
    \end{equation}
    By the augmented estimation error covariance in (\ref{E13}), if $\| P(k) \|_2 \le \| P(k+1) \|_2$, then
    \begin{equation}
    \label{E20}
    \begin{aligned}
    0 
    & \le \left( \lambda^2 -1 \right) \| P(k) \|_2  \\
    & \ \ \ \ + \eta^2 \delta_d^2 \|Q_v\|_2 + \left(1+ \eta \delta_c\right)^2 \delta_{\gamma}^2 \|Q_w\|_2
    \end{aligned}
    \end{equation}
    It turns out that
    \begin{equation}
    \label{E21}
    \| P(k) \|_2 \le \frac{\eta^2 \delta_d^2 \|Q_v\|_2 + \left(1+ \eta \delta_c\right)^2 \delta_{\gamma}^2 \|Q_w\|_2}{1-\lambda^2} := \delta_{p^1}
    \end{equation}
    If $\| P(k-1) \|_2 \le \| P(k) \|_2$, then $\| P(k-1) \|_2 \le \delta_{p^1}$ and 
    \begin{equation}
    \label{E22}
    \begin{aligned}
    \| P(k) \|_2 
    & \le \lambda^2 \delta_{p^1} + \eta^2 \delta_d^2 \|Q_v\|_2 \!+\! \left(1+ \eta \delta_c\right)^2 \delta_{\gamma}^2 \|Q_w\|_2 \\
    & := f_p(\delta_{p^1})
    \end{aligned}
    \end{equation}
    If $\| P(k-1) \|_2 \ge \| P(k)\|_2 \ge \| P(k+1)\|_2$ at all instants, then $\|P(k)\|_2 \le \|P(k-1)\|_2 \le ... \le \|P(k_0)\|_2 := \delta_{p^0}$.
    Now, we can conclude that $\| P(k) \|_2$ is bounded as
    \begin{equation}
    \label{E23}
    \| P(k) \|_2 \le \max \{\delta_{p^1}, f_p(\delta_{p^1}), \delta_{p^0}\}
    \end{equation}
    By the boundedness of $\|P(k)\|_2$, one also has that $\|P_i(k)\|_2$ is bounded. This completes the proof. \\
    \footnotesize{
    \bibliographystyle{ieeetr}

\begin{thebibliography}{99}
    	\bibitem{c1}
    	F. N. Bailey, “The application of Lyapunov’s second method to interconnected systems,” \textit{Journal of the Society for Industrial and Applied Mathematics, Series A: Control}, vol. 3, no. 3, pp. 443–462, 1965.
    	\bibitem{c2}
    	V. Kekatos and G. B. Giannakis, “Distributed robust power system state estimation,” \textit{IEEE Transactions on Power Systems}, vol. 28, no. 2, pp. 1617–1626, 2013.
    	\bibitem{c3}
    	Z. Feng, G. Hu, Y. Sun, and J. Soon, “An overview of collaborative robotic manipulation in multi-robot systems,” \textit{Annual Reviews in Control}, vol. 49, pp. 113–127, 2020.
    	\bibitem{c4}
    	M. Dickison, S. Havlin, and H. E. Stanley, “Epidemics on interconnected networks,” \textit{Physical Review. E, Statistical, Nonlinear, and Soft Matter Physics}, vol. 85, no. 6, p. 066109, 2012.
    	\bibitem{c5}
    	W. Li, Y. Jia, and J. Du, “State estimation for stochastic complex networks with switching topology,” \textit{IEEE Transactions on Automatic Control}, vol. 62, no. 12, pp. 6377–6384, 2017.
    	\bibitem{c6}
    	Y. Huang, I. Tienda-Luna, and Y. Wang, “Reverse engineering gene regulatory networks,” \textit{IEEE Signal Processing Magazine}, vol. 26, no. 1, pp. 76–97, 2009.
    	\bibitem{c7}
    	J. Lian, “Special section on control of complex networked systems (CCNS): Recent results and future trends,” \textit{Annual Reviews in Control}, vol. 47, pp. 275–277, 2019.
    	\bibitem{c8}
    	C. Kwon and I. Hwang, “Sensing-based distributed state estimation for cooperative multiagent systems,” \textit{IEEE Transactions on Automatic Control}, vol. 64, no. 6, pp. 2368–2382, 2019.
    	\bibitem{c9}
    	P. Yang, R. A. Freeman, and K. M. Lynch, “Multi-agent coordination by decentralized estimation and control,” \textit{IEEE Transactions on Automatic Control}, vol. 53, no. 11, pp. 2480–2496, 2008.
    	\bibitem{c10}
    	R. Olfati-Saber, “Distributed Kalman filtering for sensor networks,” in \textit{2007 46th IEEE Conference on Decision and Control}, (New Orleans, LA, USA), pp. 5492–5498, IEEE, Dec. 2007.
    	\bibitem{c11}
    	B. Chen, W. A. Zhang, and L. Yu, “Distributed finite-horizon fusion Kalman filtering for bandwidth and energy constrained wireless sensor networks,” \textit{IEEE Transactions on Signal Processing}, vol. 62, no. 4, pp. 797–812, 2014.
    	\bibitem{c12}
    	W. A. Zhang and L. Shi, “Sequential fusion estimation for clustered sensor networks,” \textit{Automatica}, vol. 89, pp. 358–363, 2018.
    	\bibitem{c13}
    	C. Sanders, E. Tacker, T. Linton, and R. Ling, “Specific structures for large-scale state estimation algorithms having information exchange,” \textit{IEEE Transactions on Automatic Control}, vol. 23, no. 2, pp. 255–261, 1978.
    	\bibitem{c14}
    	A. Haber and M. Verhaegen, “Moving horizon estimation for large-scale interconnected systems,” \textit{IEEE Transactions on Automatic Control}, vol. 58, no. 11, pp. 2834–2847, 2013.
    	\bibitem{c15}
    	U. A. Khan, “Distributing the Kalman filter for large-scale systems,” \textit{IEEE Transactions on Signal Processing}, vol. 56, no. 10, pp. 4919–4935, 2008.
    	\bibitem{c16}
    	S. S. Stanković, M. S. Stanković, and D. M. Stipanović, “Consensus based overlapping decentralized estimation with missing observations and communication faults,” \textit{Automatica}, vol. 45, no. 6, pp. 1397–1406, 2009.
    	\bibitem{c17}
    	B. Chen, G. Hu, D. W. C. Ho, and L. Yu, “Distributed Kalman filtering for time-varying discrete sequential systems,” \textit{Automatica}, vol. 99, pp. 228–236, 2019.
    	\bibitem{c18}
    	B. Chen, G. Hu, D. W. Ho, and L. Yu, “Distributed estimation for discrete-time interconnected systems,” in \textit{2019 Chinese Control Conference (CCC)}, (Guangzhou, China), pp. 3708–3714, IEEE, July 2019.
    	\bibitem{c19}
    	B. Chen, G. Hu, D. W. C. Ho, and L. Yu, “Distributed estimation and control for discrete time-varying interconnected systems,” \textit{IEEE Transactions on Automatic Control}, 2021. doi: 10.1109/TAC.2021.3075198.
    	\bibitem{c20}
    	Y. Zhang, B. Chen, L. Yu, and D. W. C. Ho, “Distributed Kalman filtering for interconnected dynamic systems,” \textit{IEEE Transactions on Cybernetics}, 2021. doi: 10.1109/TCYB.2021.3072198.
    	\bibitem{c21}
    	M. Farina, G. Ferrari-Trecate, and R. Scattolini, “Moving horizon partition-based state estimation of large-scale systems,” \textit{Automatica}, vol. 46, no. 5, pp. 910–918, 2010.
    	\bibitem{c22}
    	S. Riverso, M. Farina, R. Scattolini, and G. Ferrari-Trecate, “Plug-and-play distributed state estimation for linear systems,” in \textit{52nd IEEE Conference on Decision and Control}, (Firenze), pp. 4889–4894, IEEE, 2013.
    	\bibitem{c23}
    	S. Riverso, D. Rubini, and G. Ferrari-Trecate, “Distributed bounded-error state estimation based on practical robust positive invariance,” \textit{International Journal of Control}, vol. 88, no. 11, pp. 2277–2290, 2015.
    	\bibitem{c24}
    	N. Sandell, P. Varaiya, M. Athans, and M. Safonov, “Survey of decentralized control methods for large scale systems,” \textit{IEEE Transactions on Automatic Control}, vol. 23, no. 2, pp. 108–128, 1978.
    	\bibitem{c25}
    	A. N. Michel, “On the status of stability of interconnected systems,” \textit{IEEE Transactions on Systems, Man, and Cybernetics}, vol. 13, no. 4, pp. 439–453, 1983.
    	\bibitem{c26}
    	H. Ito, “A geometrical formulation to unify construction of Lyapunov functions for interconnected iISS systems,” \textit{Annual Reviews in Control}, vol. 48, pp. 195–208, 2019.
    	\bibitem{c27}
    	A. N. Michel, “Stability analysis of interconnected systems,” \textit{SIAM Journal on Control}, vol. 12, no. 3, pp. 554–579, 1974.
    	\bibitem{c28}
    	W. M. Haddad and S. G. Nersesov, \textit{Stability and Control of Large-Scale Dynamical Systems: A Vector Dissipative Systems Approach}. Princeton: Princeton University Press, 2011.
    	\bibitem{c29}
    	S. N. Dashkovskiy, B. S. R¨uffer, and F. R. Wirth, “Small gain theorems for large scale systems and construction of ISS Lyapunov functions,” \textit{SIAM Journal on Control and Optimization}, vol. 48, no. 6, pp. 4089–4118, 2010.
    	\bibitem{c30}
    	H. Ito, “State-dependent scaling problems and stability of interconnected iISS and ISS systems,” \textit{IEEE Transactions on Automatic Control}, vol. 51, no. 10, pp. 1626–1643, 2006.
    	\bibitem{c31}
    	P. Moylan and D. Hill, “Stability criteria for large-scale systems,” \textit{IEEE Transactions on Automatic Control}, vol. 23, no. 2, pp. 143–149, 1978.
    	\bibitem{c32}
    	M. Vidyasagar, ed., \textit{Input-output analysis of large-scale interconnected systems. Decomposition, well-posedness and stability}. Berlin/Heidelberg: Springer-Verlag, 1981.
    	\bibitem{c33}
    	E. Agarwal, S. Sivaranjani, V. Gupta, and P. J. Antsaklis, “Distributed synthesis of local controllers for networked systems with arbitrary interconnection topologies,” \textit{IEEE Transactions on Automatic Control}, vol. 66, no. 2, pp. 683–698, 2021.
    	\bibitem{c34}
    	A. A. Alam, A. Gattami, and K. H. Johansson, “An experimental study on the fuel reduction potential of heavy duty vehicle platooning,” in \textit{13th International IEEE Conference on Intelligent Transportation Systems}, pp. 306–311, Sept. 2010.
    	\bibitem{c35}
    	J. Yang, W. A. Zhang, and F. Guo, “Dynamic state estimation for power networks by distributed unscented information filter,” \textit{IEEE Transactions on Smart Grid}, vol. 11, no. 3, pp. 2162–2171, 2020.
    	\bibitem{c36}
    	S. Boyd, L. El Ghaoui, E. Feron, and V. Balakrishnan, eds., \textit{Linear Matrix Inequalities in System and Control Theory}. Philadelphia, PA, USA: Society for Industrial and Applied Mathematics, 1994.
    \end{thebibliography}
    }
\end{document}